\documentclass[12pt,a4paper]{article}
\usepackage{graphicx}

\begin{document}
\begin{center}
{\LARGE \bf Entanglement in a Quantum Mixed-Spin Chain}

\vskip0.8cm {\it Xiang Hao and Shiqun Zhu*}

\vskip0.2cm
{\it China Center of Advanced Science and Technology(World Laboratory),\\
P. O. Box 8730, Beijing 100080, People's Republic of China\\
and\\
School of Physical Science and Technology, Suzhou University,\\
Suzhou, Jiangsu 215006, People's Republic of China**}
\end{center}

\begin{abstract}
The entanglement in a general mixed spin chain with arbitrary spin
$S$ and $1/2$ is investigated in the thermodynamical limit. The
entanglement is witnessed by the magnetic susceptibility which
decides a characteristic temperature for an entangled thermal
state. The characteristic temperature is nearly proportional to
the interaction $J$ and the mixed spin $S$. The bound of
negativity is obtained on the basis of the magnetic
susceptibility. It is found that the macroscopic magnetic
properties are affected by the quantum entanglement in the real
solids. Meanwhile, the entanglement can be quantitatively
evaluated by the thermodynamical observable.
\end{abstract}


* Corresponding author, email: szhu@suda.edu.cn

** Mailing address
\newpage

The quantum entanglement can exhibit the quantum non-locality that
cannot be explained by the classic laws. The entanglement is very
essential in the quantum computation and quantum information
processing [1,2]. As a crucial resource, the entanglement has been
investigated by some basic measures [3-5] in many systems [6-9]. The
entanglement of formation [3] and the relative entropy of
entanglement [4] are used for two qubits. The negativity [5] is used
to qualify the entanglement for any bipartite system.

The entanglement consists not only in microscopic quantum systems
[6,7], but also in macroscopic systems [8-10]. It is demonstrated
that the macroscopic thermodynamical observable contains the
entanglement. From the separability criterion [11], the entanglement
witness [12-19] has been proposed to evaluate the entanglement in
the real solids. An entanglement witness is an observable which has
an positive expected value for a separable (unentangled) state, a
negative value for an entangled state. The effect of the
entanglement in solids has been observed in the experimental
situation [10]. The thermal energy [13-16], the heat capacity [19],
and the magnetic susceptibility [10, 20-22] have been considered as
the witness for the quantum spin-$\frac 12$ systems. They decide a
characteristic temperature below which an entangled thermal state is
obtained. Besides the general spin-$\frac 12$ system [23], the
properties of the quantum mixed or alternating spin system [24, 25]
is also attracting due to the peculiar phenomenon of spin-Peierls
transition. For an inorganic compound, the quasi one-dimensional
bimetallic chains $ACu(pbaOH)(H_2O_3)\cdot 2H_2O$, with each unit
cell containing two different spins ($S, 1/2$), have been
synthesized [25]. The compounds of ACu may be considered as a
Heisenberg mixed spin ($S, 1/2$) chain with antiferromagnetic
interactions. It is necessary to study the entanglement in the real
mixed spin systems.

In this Letter, the entanglement in a general mixed spin chain is
investigated on the basis of the experimental measurements of the
magnetic susceptibility. The entanglement is witnessed by the
magnetic susceptibility. The characteristic temperatures for the
existing entanglement are determined by the entanglement witness.
The experimental measurements of the magnetic susceptibility
published previously are fitted by the model of one-dimensional
Heisenberg mixed spin chain. The bound of negativity is obtained
in the thermodynamical limit.

For an isotropic mixed spin ($S, 1/2$) chain, the Hamiltonian $H$
can be expressed by
\begin{equation}
\label{eq:1}
 \mathcal{H}=\sum_{i=1}^{n/2} \mathcal{H}_{2i}
\end{equation}
where $\mathcal{H}_{2i}=J(\vec{S}_{2i-1}\cdot\vec{s}_{2i}
 +\vec{s}_{2i}\cdot\vec{S}_{2i+1})$, $\vec{S}$ and $\vec{s}$ are the spin vector of
arbitrary spin $S$ and $1/2$ respectively. The number of spins is
$n$, and the periodic condition of $n+1=1$ is assumed. The
exchange interaction of $J>0$ and $J<0$ denote the
antiferromagnetic and ferromagnetic case respectively. In the
following discussion, an antiferromagnetic chain is considered.

The state at a thermal equilibrium temperature $T$ is
$\rho(T)=e^{-H/k_BT}/Z$ where $Z$ is the partition function and
$k_B$ is Boltzmann constant. The general expression for the
magnetic susceptibility at zero-field can be written as
\begin{equation}
\label{eq:2} \chi_{\alpha}=\chi= \frac {g^2\mu_B^2}{k_BT}
\sum_{i,j=1}^n \mathcal{G}^{\alpha}_{ij}, (\alpha=x,y,z)
\end{equation}
where $\chi$ is the magnetic susceptibility at the thermal
equilibrium temperature $T$, $g$ is the g-factor, $\mu_B$ is the
Bohr magneton, and $\mathcal{G}^{\alpha}_{ij}$ represents the
correlation between the site $i$ spin and $j$ spin along $\alpha$
direction. It is assumed that the correlations of two spins which
are not nearest neighboring are small enough to be ignored. This
implies that the magnetic susceptibility can be approximately
obtained by $\chi \simeq \frac {g^2\mu_B^2n}{k_BT}[\frac 18+\frac
{S^2}2+\frac {\mathcal{G}_1}3]$, where $\mathcal{G}_1$ is the
correlation between two nearest neighboring spins. In the isotropic
mixed spin chains, any two nearest-neighboring thermal state is
SU(2)-invariant [26]. If the separability principle [11]is followed,
$\mathcal{G}_1<- \frac S2$ will be satisfied for an entangled
thermal state. In the thermodynamical limit, the entanglement
witness $\mathcal{W}$ can be given by
\begin{equation}
\label{eq:3} \mathcal{W}=\chi-\frac
{g^2\mu_B^2n(12S^2-4S+3)}{24k_BT}
\end{equation}
If the value $\mathcal{W}$ of the witness is negative, the thermal
state is entangled. The separable state is for the value of
$\mathcal{W}>0$. Then Eq.(3) is the generalized form of the
entanglement witness for the one-dimensional mixed spin ($S,1/2$)
chain. From Eq.(3), the presence of the entanglement will be
detected if the macroscopic susceptibility is measured and the
parameters of the structure, i. e. the values of $g$ and $n$, are
known. There is a characteristic temperature $T_c$ for
$\mathcal{W}=0$. Below the temperature, an entangled thermal state
is attained in the spin system. The characteristic temperature $T_c$
is plotted as a function of the mixed spin length $S$ and
interaction $J$ in Fig.1. Here the height of the bar denotes the
value of $T_c$. From Fig.1, it is seen that the values of $T_c$ are
increased with the increase of the exchange interaction $J$. It is
also clear that $T_c$ is almost linearly increased with the spin $S$
for the same interaction. By the numerical calculation, the
characteristic temperature is approximately expressed as $T_c\approx
J(a_0S+b_0)/k_B$ where $a_0$ and $b_0$ are two constants determined
by the property of the material. The values of $T_c$ can be improved
in the condition of high mixed spin and the strong exchange
interaction.
\begin{table}
\caption{\label{tab:temperatures}The characteristic temperatures
for compounds*.}

\begin{tabular}{cccc}
compounds &  $J$ ($cm^{-1}$) &  spin ($S, 1/2$) &  approx. $T_c$ ($K$)\\
\hline CN &  5.12$k_B$& ($1/2, 1/2$)
&4.7\\
ACu (A=Ni) &  81.4  & ($1, 1/2$) & 125\\
ACu (A=Co) &  18  & ($3/2, 1/2$) & 26\\
ACu (A=Fe) &  20  & ($2, 1/2$) & 32\\
ACu (A=Mn) &  23.44  & ($5/2, 1/2$) & 40\\

\end{tabular}

*The data of CN are from [20] and others are from
[25].
\end{table}
For an inorganic compound, the bimetallic chain ACu [25], (A=Mn, Fe,
Co, Ni) has been synthesized. At a finite temperature, the
macroscopic properties can be explained by the one-dimensional mixed
spin chain. Using Eq.(3) and the measurements of the susceptibility
[20, 25], the characteristic temperature $T_c$ for the entanglement
is approximately given in Table 1. The special compound of CN ($1/2,
1/2$) is also evaluated. The value of $T_c$ is about $4.7K$ which is
in good agreement with that given by [20]. The ACu compound
corresponds to the Heisenberg spin ($S^A, 1/2$) chain with $S^A=5/2,
2, 3/2, 1$. The presence of the macroscopic entanglement is
qualitatively described by the entanglement witness.

Although the entanglement witness $\mathcal{W}$ indicates the
presence of the entanglement, it is still necessary to
quantitatively study the relation of the quantum entanglement and
the macroscopic thermodynamical observable. From the separability
principle, the negativity [5] for the entanglement measurement can
be introduced by
\begin{equation}
\label{eq:4}
 \mathcal{N}(\rho)=|\sum_i\tau_i|
\end{equation}
where $\tau_i$ is the $i$th negative eigenvalue of $\rho^{T}$
which is the partial transpose of the mixed state $\rho$. The
measure corresponds to the absolute value of the sum of negative
eigenvalues of $\rho^{T}$. The state $\rho$ is unentangled if all
eigenvalues of $\rho^{T}$ are nonnegative. For an isotropic mixed
spin ($S, 1/2$) chain, thermal states between any two nearest
neighboring spins $\rho_{2i}$ is SU(2)-invariant. The possible
negative eigenvalue of $\rho_{2i}^{T}$ can be written as
\begin{equation}
\label{eq:5} \tau=\frac
{S+2\mathcal{G}_1}{\mathcal{D}(\mathcal{D}-1)}
\end{equation}
where the degree of degeneracy is $2S$, and $\mathcal{D}=2S+1$
represents the dimension of the mixed spin $\vec{S}$. Considering
the effect of the correlation between any two spins, the lower
bound of the negativity is obtained
\begin{equation}
\label{eq:6} \mathcal{N}(\rho)> -\frac
{6\mathcal{W}k_BT}{\mathcal{D}g^2\mu_B^2n}
\end{equation}
where $\mathcal{N}(\rho)$ denotes the entanglement between two
nearest neighboring spins in the thermodynamical limit. From Eq.
(6), it is seen that the entanglement exists if Eq.(6) holds with
$\mathcal{N}(\rho)>0$. The Eq.(6) is generalized to the isotropic
mixed spin chain with the arbitrary spin of $S$. It is seen that the
bound of negativity is proportional to the value of $-\mathcal{W}$.
Meanwhile, the entanglement is decreased with the increase of the
macroscopic value $\chi T$. The entanglement can be quantitatively
evaluated in some degree when the magnetic susceptibility is
measured. The bound of the entanglement may be decided by the
measurements of the magnetic susceptibility. It is seen that the
entanglement $\mathcal{N}(\rho)$ can affect the macroscopic magnetic
properties.

For the special case of $S=1/2$, the copper nitrate Cu-HTS [25] is
approximately regarded as a one-dimensional antiferromagnetic
Heisenberg chain. According to the experimental measurements of the
magnetic susceptibility [24], the measured data of $\chi$ are shown
by the circles in Fig. 2(a). When $T\rightarrow0$, the magnetic
susceptibility is finite. With the increase of the temperature, the
values of $\chi$ is increased to a maximum at about $T=20K$. Then
the magnetic susceptibility declines as the temperature $T$ is
increased. In Fig. 2(a), the experimental data are fitted by the
theoretical curve when the interaction of $J=10.2cm^{-1}$ and the
g-factor of $g=2.06$. The solid line is the theoretical curve of the
Heisenberg linear chain. The theoretical curve is in good agreement
with the measured magnetic susceptibility. Above the temperature of
about $T=12K$, the effects of the correlation between
non-nearest-neighboring spins can be reasonably negligible. This is
the reason that the effects between non-nearest-neighboring spins
exponentially decrease as the separation of the sites is increased.
However, at the low temperatures, these effects need to be
considered. Using Eq.(6) and the measurements of the susceptibility,
the experimental data of the bound of negativity is plotted by the
circles in Fig. 2(b). The entanglement drops rapidly with the
temperature, and vanishes at about $T=33K$. For a pair of spins, the
correlation function is theoretically expressed as
$\mathcal{G}_1=-3(1-e^{-J/k_BT})/4(1+3e^{-J/k_BT})$. According to
Eq.(4), the theoretical prediction of the negativity is plotted by
the solid line in Fig. 2(b). Below $T=12K$, the theoretical value is
slightly higher. It is obvious that the theoretical curve obtained
from Eq.(6) is in good agreement with the measurement of the
susceptibility. To some extent, the quantum entanglement is
quantitatively evaluated in the real solids.

The general case of the mixed spin with $S\neq \frac 12$ needs also
to be investigated. As an example, the bimetallic compound chain of
NiCu [25] can be considered as the mixed spin ($1, 1/2$) model.
Above a certain temperature, the magnetic properties can be
explained by the one-dimensional mixed spin chain. Due to the
effects of three-dimensional and the long range exchange
interaction, the magnetic susceptibility cannot be fitted using the
one-dimensional linear chain at low temperatures. Above a certain
temperature, the spin model can be simplified as the one-dimensional
chain with the nearest neighboring interaction. Therefore, the
macroscopic quantity of $\chi T$ and the theoretical prediction are
shown in Fig. 3(a) at the temperature range of $25-250K$. The
circles denote the experimental data of $\chi T$ [25]. Above the
temperature
 of $T=8K$, $\chi T$ is declined to the minimum at about
$T=75K$. The solid line is the theoretical fit by the
one-dimensional mixed spin chain. It is seen that the values of
$\chi T$ is in good agreement with the experimental data above
$T=25K$. Similarly, the bound of the negativity is illustrated in
Fig. 3(b). The results of the experimental data are shown by the
circles. From the measured data in Fig. 3(b), it is clear that the
entanglement monotonously declines with the increase of the
temperature. It is noted that the correlations of the
non-nearest-neighboring spins will decrease with the temperature.
Above the temperature $T=80K$, the effect between two nearest
neighboring spins $(1, 1/2)$ dominates the macroscopic
susceptibility. But the contributions from non-nearest-neighboring
spins, i. e. $(1,1)$ and $(1/2,1/2)$, need still to be considered at
the range of $T<75K$. Thus, the correction item needs to be added to
Eq.(6). By the numerical calculation, the additional item is
proportional to $\mathcal{P}\langle \vec{S}_{2i-1}\cdot\vec{s}_{2i}
\rangle$, where $\mathcal{P}$ is the polynomial decreasing function
of the temperature. For the case of NiCu, the function can be
expressed as $\mathcal{P}\sim 0.11(J/k_BT)-0.07(J/k_BT)^2$. Above
the temperature $T=122k$, there is no entanglement in the compound.
For low temperatures of $T<25K$, the spin model cannot be regarded
as a one-dimensional linear chain with the antiferromagnetic
interaction. Therefore, only the entanglement above a certain
temperature can be calculated by Eq.(6). For a pair of spins, the
correlation is given by $\mathcal{G}_1=-5(1-e^{-3J/2k_B
T})/[6(1+2e^{-3J/2k_B T})]$. From Eq.(4), the theoretical value of
the negativity is plotted by the solid line. It is seen that the
theoretical curve is slightly higher than the results of the
susceptibility measurements when the temperature is low.

From the two kinds of solid compounds, it is demonstrated that the
quantum entanglement plays an essential role in the macroscopic
phenomena. The magnetic susceptibility embodies the quantum
entanglement. From the experimental measurements of the
susceptibility, the entanglement is quantitatively evaluated. It
is likely that the quantum entanglement can be used as a resource
of the quantum information processing in the thermodynamical
limit.

In summary, the entanglement in a general mixed spin ($S, 1/2$)
chain is investigated using the macroscopic magnetic
susceptibility. The entanglement witness for arbitrary spin $S$ is
expressed by the susceptibility. The entanglement witness
determines a characteristic temperature $T_c$ below which an
entangled thermal state can be obtained. It is seen that the
values of $T_c$ are nearly proportional to the exchange
interaction $J$ and the mixed spin $S$. The characteristic
temperatures for the ACu compounds are approximately determined.
It is likely that the entanglement can be detected at a rather
high temperature in the solids with high spin and the strong
interaction. The relation of the quantum entanglement and the
macroscopic susceptibility is deduced. The bound of negativity can
be obtained by the experimental measurements of the
susceptibility. To some extent, the entanglement can be
quantitatively evaluated by the susceptibility. It is demonstrated
that the macroscopic magnetic properties can be affected by the
quantum entanglement in the real solids.

It is a pleasure to thank Yinsheng Ling, Jianxing Fang, and Qing
Jiang for their many fruitful discussions about the topic.

\newpage

{\Large \bf Reference}

1. D. P. DiVincenzo and D. Eacon and J. Kempe and G. Burkard and K.
B. Whaley, Nature(London)408, 339(2000).

2. A. Vaziri and G. Weihs and A. Zeilinger, Phys. Rev. Lett.89,
240401(2002).

3. W. K. Wootters, Phys. Rev. Lett.80, 2245(1998).

4. V. Vedral and M. B. Plenio, Phys. Rev. A57, 1619(1998).

5. G. Vidal and R. F. Werner, Phys. Rev. A65, 032314(2002).

6. M. C. Arnesen and S. Bose and V. Vedral, Phys. Rev. Lett.87,
017901(2001).

7. G. L. Kamta and A. F. Starace, Phys. Rev. Lett.88, 107901(2002).

8. S. Gu and S. Deng and Y. Li and H. Lin, Phys. Rev. Lett.93,
086402(2004).

9. C. Lunkes and C. Brukner and V. Vedral, Phys. Rev. Lett.95,
030503(2005).

10. S. Ghosh and T. F. Rosenbaum and G. Aeppli and S. Coppersmith,
Nature(London)425, 48(2003).

11. A. Peres, Phys. Rev. Lett.77, 1413(1996).

12. X. Wang and P. Zanardi, Phys. Lett. A301, 1(2002).

13. M. R. Dowling and A. C. Doherty and S. D. Bartlett, Phys. Rev.
A70, 062113(2004).

14. K. Maruyama and F. Morikoshi and V. Vedral, Phys. Rev. A71,
012108(2005).

15. G. Toth, Phys. Rev. A71, 010301(R)(2005).

16. P. Hyllus and O. Guhne and D. Bruss and M. Lewenstein, Phys.
Rev. A72, 012321(2005).

17. M. Lewenstein and B. Kraus and J. I. Cirac and P. Horodecki,
Phys. Rev. A62, 052310(2000).

18. A. Sanpera and D. Bruss and M. Lewenstein, Phys. Rev. A63,
050301(R)(2001).

19. M. Wiesniak and V. Vedral and C. Brukner, e-print
quant-ph/0508193.

20. C. Brukner and V. Vedral, and A. Zeilinger, e-print
quant-ph/0410138.

21. M. Wiesniak and V. Vedral and C. Brukner, e-print
quant-ph/0503037.

22. T. Vertesi and E. Bene, e-print quant-ph/0503726.

23. Q. Jiang and Z. Y. Li, Phys. Rev. B40, 11264(1989).

24. W. E. Hatfield, J. Appl. Phys.52, 1985(1981).

25. P. J. van Koningsbruggen and O. Kahn and K. Nakatani and Y. Pei
and J. P. Renard and M. Drillon and P. Legoll, Inorg. Chem.29,
3325(1990).

26. J. Schliemann, Phys. Rev. A68, 012309(2003).

\newpage

{\Large \bf Figure Captions}

{\bf Fig. 1.}

The characteristic temperature $T_c$ is plotted when the mixed spin
$S$ and the exchange interaction $J$ are varied. The height of the
bar denotes the value of $T_c$. It is linearly increased with the
increasing values of $S$ and $J$.

{\bf Fig. 2.}

(a). The magnetic susceptibility of the compound Cu-HTS is plotted
as a function of the temperature $T$. The circles are the measured
data [24]. The parameters are taken from Ref. [24] with the
interaction $J=10.2cm^{-1}$ and the g-factor $g=2.06$. The
theoretical prediction is plotted by the solid line.

(b). Based on the experimental measurements, the bound of the
negativity is calculated and represented by the circles. The solid
line is the theoretical curve.

{\bf Fig. 3.}

(a). The measurements of the magnetic susceptibility for $NiCu$ from
Ref. [25] are denoted by the circles. Above the temperature $T=75K$,
the magnetic properties can be explained by the one-dimensional
mixed spin chain. The solid line is the theoretical fit with the
interaction $J=81.4cm^{-1}$ and the g-factor $g=2.15$.

(b). From the experimental data of $\chi T$, the bound of the
negativity is plotted as the circles. The solid line is the
theoretical fit.

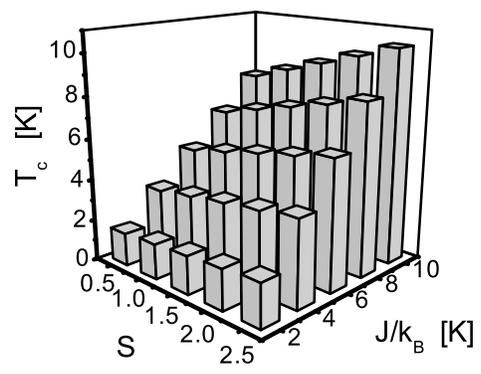
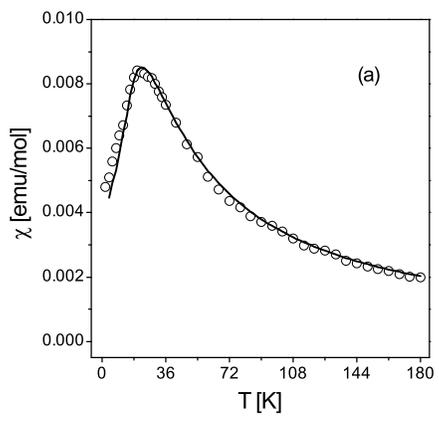
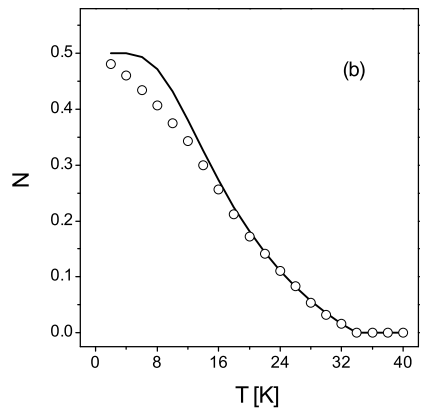
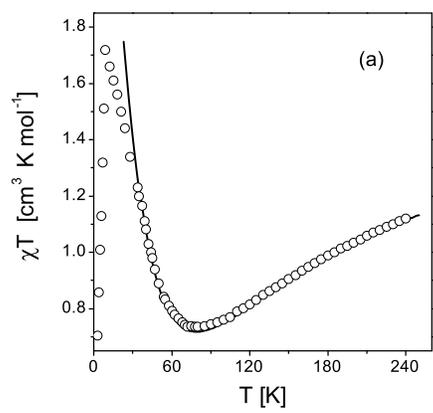
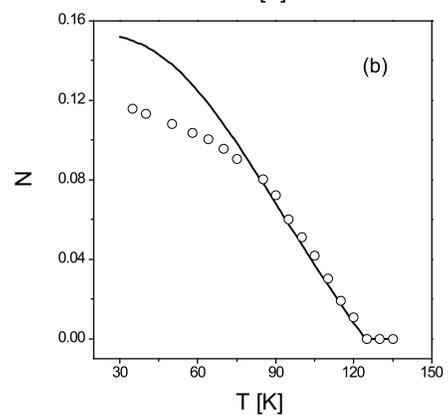
\end{document}